\documentstyle[12pt]{article}
\normalbaselines
\begin{document}
\bibliographystyle{unsrt}

\vbox{\vspace{38mm}}
\begin{center}
{\LARGE \bf THE KP EQUATION FROM  PLEBANSKI \\[2mm]
AND $SU(\infty)$ SELF-DUAL YANG-MILLS   }\\[5 mm]

Carlos Castro \\{I.A.E.C 1407 Alegria\\
Austin, Texas 78757  USA }\\[3mm]
(July 1993)\\[5mm]
\end{center}

\begin{abstract}
     Starting from a self-dual $SU(\infty)$ Yang-Mills theory in $(2+2)$
dimensions, the Plebanski second heavenly equation is obtained after a
suitable dimensional reduction. The self-dual gravitational background is
the cotangent space of the internal two-dimensional Riemannian surface
required in the formulation of $SU(\infty)$ Yang-Mills theory. A subsequent
dimensional reduction leads to the KP equation in $(1+2)$ dimensions after
the relationship from the Plebanski second heavenly function, $\Omega$, to
the KP function, $u$, is obtained. Also a complexified KP equation
is found when a different dimensional reduction scheme is performed
.  Such relationship between $\Omega$ and $u$  is based on the
correspondence between the $SL(2,R)$ self-duality conditions in $(3+3)$
dimensions of Das, Khviengia, Sezgin (DKS) and the ones of $SU(\infty)$
 in $(2+2)$ dimensions . The generalization to the
Supersymmetric KP equation  should be straightforward by
extending the construction of the bosonic case to the previous Super-Plebanski
equation, found by us in
[1], yielding  self-dual supergravity backgrounds
in terms of the light-cone  chiral superfield, $\Theta$, which is the
supersymmetric analog of $\Omega$. The most important consequence of this
Plebanski-KP correspondence is that $W$ gravity can be seen as the gauge
theory of  $\phi$-diffeomorphisms  in the space of dimensionally-reduced
$D=2+2,~SU^*(\infty)$ Yang-Mills instantons. These $\phi$ diffeomorphisms
preserve a volume-three-form and are, precisely, the ones which provide the
Plebanski-KP correspondence.
\end{abstract}

\section {Introduction}
   The infinite dimensional Lie algebra of area preserving diffeomorphisms
of a surface, $sdiff~\Sigma$, plays a fundamental role in the physics of
membranes; in the connection between gauge theories and strings; large $N$
models calculations, quantum groups, integrable models, $W_{\infty}$ algebras;
to name a few. In a previous paper [1]  we were able to show that
$SU(\infty)$ self-dual Yang-Mills equations on a four-dimensional flat
Euclidean background, were equivalent to Plebanski's second heavenly
equation for the self-dual gravitational background, which is associated
with the cotangent space of a suitable Riemannian surface, $T^*\Sigma$. Such
surface in question was the
internal two-dimensional surface required in the formulation of
$SU(\infty)$ Yang-Mills theories by Floratos et al [2]. Such equivalence
only occured after a suitable dimensional reduction has taken place. The
results
were generalized to the supersymmetric case where, for the first time as far
as we know, we obtained the supersymmetric analog of Plebanski's  second
heavenly equation [1]. In a subsequent paper [3] , we were also able to show,
generalizing Park's results for the bosonic case [4], that the $N=2$ SWZNW
model, valued in the area-preserving diffeomorphisms group of $\Sigma$, was
equivalent to self-dual supergravity in four dimensions. The four
dimensional manifold was comprised of suitable coordinate- patches obtained by
gluing
pieces of $\Sigma$ with those of $\cal M$, the base two-dimensional manifold
where the $N=2$ non-linear sigma model was defined.

    A further Killing symmetry reduction of the (super)Plebanski (first
heavenly equation) yields the $SL(\infty)$ continual (super)Toda field
equations in three dimensions. The symmetry algebra of the
$SL(\infty)$ continual (super) Toda equations was, precisely, the classical
(super) $W_\infty$ algebra, and this can be obtained as a  Killing symmetry
reduction of the symmetry algebra of the (super) Plebanski equation; i.e.
a (super) $CP^1$-extension of $sdiff~\Sigma$. It has been known for some
time that various conformal algebras in two dimensions ($W_N$ algebras,
etc..) arise as Hamiltonian structures (Poisson brackets) of integrable
systems [5]. In fact, the first Hamiltonian structure of the
$(2+1)$-dimensional $KP$ hierachy, can be identified with the classical
$W_{1+\infty}$ algebra which
has been the subject of paramount interest in recent years [6].

    The KP equation is the subject of this work. Das, Khviengia and Sezgin
(DKS) [7] showed, contrary to people's beliefs, that the KP equation could
be obtained as a self-duality condition for $sl(2,R)$- valued Yang-Mills on
a $(3+3)$ dimensional paracomplex manifold. A suitable dimensional
reduction $ and$ an ansatz was required,  as it is the case for all of these
models obtained from a self-duality condition.  In previous work [8] we
showed that the self-dual supermembrane in $(4+1)$-dimensions  was an
integrable system equivalent to the super Toda molecule for
$SU(N\rightarrow \infty)$ and for the $minimal$ embedding of $SU(2)$ into
$SU(N)$. The results for the bosonic case were found by [9]. Since
the Lie-algebra   $sl(2,R)$ is locally isomorphic to $su(1,1)$ and
$su(2)$ is locally isomorphic to $sl(1,H)$, we can can use the result
 that $sl(N,H)\sim su^*(2N)$ to embed $sl(2,R)$ into $sl(N,H)$
and, subsequently, into
$SU^*(N\rightarrow \infty)$. Essentially this can be done because a linear
combination of
the Pauli
spin matrices $obey$ the $sl(2,R)$ algebra. In this $N\rightarrow \infty$
limit, it is then  natural to ask if the KP equation can be
obtained as a self-duality condition for $SU(\infty)$ Yang-Mills in $(2+2)$
dimensions. The answer is yes.

   A derivation of the KP equation based on the asymptotic $h\rightarrow 0$
limit of the continual $sl(N+1,C)$ Toda molecule equation was given earlier by
Chakravarty and Ablowitz [10]. The continual Toda molecule equation was
obtained,  first,  by a suitable ansatz, dimensional reduction and continuous
version of the the Cartan basis for $sl(N+1,C)$. Our results here $differ$
from those of [10] by the following :

1- The method is $different$ : The derivation is based on a self-duality
condition for the $SU(\infty)$ gauge group.  2- No asymptotic nor
perturbation expansions are performed. 3- The physical role that the self-dual
(super) membrane and
the $sdiff~\Sigma$ Lie-algebra  hierarchy has on the KP hierarchy is
manifest. 4- A geometrical setting and the role played by self-dual (super)
gravity
in the derivation of the (super) KP equation is unravelled; i.e. The geometry
of
$T^*\Sigma$ has a fundamental place. Witten [11] has discussed the role that
$T^*\Sigma$ plays in the geometrical meaning of $W$ gravity  based on
Hitchin's  monopole- bundle constructions. 5- A supersymmetric extension is
straightforward by borrowing the results in [1]. 6- The
underlying higher-dimensional- than-four origin of the KP equation is shown
when one exploits  the correspondence  between the
six-dimensional self-duality condition for $SL(2,R)$ valued gauge fields
(DKS construction), and the effective six dimensional $SU(\infty)$ Yang-Mills
theory,  after the
Lie-algebra-valued potentials  are replaced
by $c$-number functions of two extra variables.

   We believe that the six reasons above should be sufficient to interest
the reader.

Therefore, the KP equation can be obtained
from a dimensional reduction of Plebanski's second heavenly equation
( valid also for the $2+2$ signature) reinforcing, evenfurther, the role of
$W_\infty$ symmetry algebras in these
integrable systems. The clue rests on the fact that we are able to embed
$sl(2,R)$ into $su(\infty)$ and on the role played by the correspondence
between DKS and $SU(\infty)$ Yang-Mills. Ultimately, everything boils down
to integrablity of the
Toda system. Bogoyavlenski  [12] has shown that the Hamiltonian for
the periodic Toda lattice looks like Einstein dynamical systems in the
theory of cosmological models; the connection between  Einstein
gravity and the KP equation is more transparent in this case . The connection
between a $basis$-dependent limit of
$SU(N\rightarrow\infty)$ and $sdiff~S^2$ was provided some time ago by
Hoppe [13]. See [13] for further details. After this lengthy introduction
we embark into explaining how we obtain the KP equation.

\section {The KP equation from Plebanski's equation}

     Foratos et al [2] were able to formulate the $N\rightarrow \infty$
limit of a $SU(N)$ Yang-Mills theory by replacing the Lie-algebra- valued
space-time dependent gauge fields, $A^a_\mu$, by $c$-number functions of
two extra bosonic coordinates parametrizing an internal two dimensional
surface (a sphere, per example) sitting over each spacetime point. In this
limit, the $SU(\infty)$ gauge theory was equivalent to a new type of gauge
principle, where gauge transformations were replaced by the $sdiff~\Sigma$
Lie-algebra and
the Lie bracket was replaced by Poisson brackets with respect to the two
coordinates, $q,p$
parametrizing the internal surface,  and the group trace was replaced by an
integration with respect to $q,p$ :

$$[A^a_\mu T_a, A^b_\nu T_b]\rightarrow \{A_\mu,A_\nu\}.\eqno(1a)$$
$$F_{\mu\nu}\rightarrow
\partial_\mu A_\nu -\partial_\nu A_\mu +\{A_\mu,A_\nu\}.\eqno (1b)$$

$$A^a_\mu (X^i)T_a =A_\mu(X^i) \rightarrow A_\mu
(X^i;q,p).~~Tr(T^aT^b)\rightarrow
\int~dqdp.\eqno (2)$$

    It is  $precisely$ the above correspondence, eqs-(1,2), which will
$provide$
for us the ansatz which shall furnish the KP equation from Plebanski's
second heavenly equation in $2+2$ dimensions. To achieve this we just need to
borrow
from the results of DKS [7] and establish a correspondence (a dictionary)
between the DKS equations and the equations (1-5) given by us in [1].
We could have presented the following ansatz, below, relating $\Omega$ to $u$.
However one would have not
known what was the underlying reason behind it and why it works. It is not
enough to write down a suitable and judicious guess for an ansatz and claim
that it is correct because it happens to work out. It is more important to
explain $why$ it works and $where$ it came from. Therefore, it is the
correspondence in eqs-(1,2) that  $explains$ why the ansatz, below,  works,
as we shall see.

    Let us choose complex coordinates for the complexified-
spacetime, $C^4$;  $y=(1/\sqrt 2)(x_1+ix_2);\tilde
y=(1/\sqrt2)(x_1-ix_2);\tilde z=(1/\sqrt 2)(x_3+ix_4)$ and $z=(1/\sqrt
2)(x_3-ix_4)$. The metric of signature $(4,0)$ and $(2+2)$
is, respectively, $ds^2=dyd\tilde y+(-)dzd\tilde z$ and the
complexified-spacetime  SDYM equations are
$F_{yz}=F_{\tilde y\tilde z} =0$ and $F_{y\tilde y}+(-)F_{z\tilde z}=0$.
The internal coordinates, $q,p$ can be incorporated into a pair of
$complex$-valued, canonical-conjugate variables ; $\hat q= Q(q,p).
{}~\hat p= P(p,q)$ such as $\{\hat q,\hat p\}_{qp}=1$. $Q,P$ are independent
maps from a sphere ( $S^2\sim CP^1$), per example, to $C^1$, such as
$Q\neq \lambda P;~\lambda =constant $.  This is in agreement with the
fact that the true symmetry algebra of Plebanski's equation is the $CP^1$
extension of the $sdiff~\Sigma$ Lie-algebra as discussed by Park [4]. We
shall relegate a  further discussion on this issue to the end of this section.

    It was the suitable dimensional reduction :
$\partial_y=\partial_{\hat q};~ -\partial_{\tilde y}=\partial_{\hat p}$
and the ansatz (where for convenience we drop the ``hats'' over the $q,p$
variables ) :

$$\partial_zA_{\tilde z}=(1/2\kappa^2)\Omega,_{zq};
\partial_{\tilde z}A_z=(1/2\kappa^2)\Omega,_{\tilde z  p};\eqno(3a)$$
$$\partial_pA_{z}=(1/2\kappa^2)\Omega,_{pp};
\partial_qA_{\tilde z}=(1/2\kappa^2)\Omega,_{qq};\eqno(3b)$$
$$\partial_qA_{z}=(1/2\kappa^2)\Omega,_{pq};
\partial_pA_{\tilde z}=(1/2\kappa^2)\Omega,_{pq};\eqno(3c)$$
$$A_y=(1/2\kappa^2)\Omega,_q ;~~~A_{\tilde y} =-(1/2\kappa^2)\Omega,_p.
\eqno(3d)$$
that yields the Plebanski equation in [1] .  $\kappa$ is a constant that has
dimensions of length and can be set to unity. The semicolon stands for partial
derivatives and $\Omega(z,\tilde z;\hat q,\hat p) $ is the   Plebanski's second
heavenly function. Upon such an anstaz and dimensional reduction, Plebanski's
second
heavenly equation was obtained in [1] :
$$(\Omega,_{\hat p\hat q})^2 -\Omega,_{\hat p\hat p}\Omega,_{\hat q\hat q}
+\Omega,_{z\hat q}
-\Omega,_{\tilde z  \hat p} =0. \eqno(4)$$

Eq-(4) yields self-dual solutions to the complexified-Einstein's equations,
and gives rise to hyper-Kahler metrics on the complexification of $T^*\Sigma$,
through a continuous self-dual deformation, represented by $\Omega$, of
the flat  metric in $(T^*\Sigma)^c$ [1].

  One of the plausible  first steps  in the dimensional-reduction of
Plebanski's equation is to
take a real-slice. A natural $real$ slice can be taken
by setting : $\tilde z=\bar z.~\tilde y=\bar y$ which implies, after using
: $\partial_y=\partial_{\hat q} ;~-\partial_{\tilde y}=\partial_{\hat p}$, that
$-(\partial_{\hat q})^*=\partial_{\hat p}$ and, hence, the Poisson-bracket
degenerates to zero; i.e. it ``collapses'' :
The quantity :$\{Q,P\}_{q,p}=\{Q,-Q^*\}_{q,p}$, if real, cannot be equal to $1$
but is
zero as one can verify by taking complex-conjugates on both sides of the
equation.  Therefore, since  the Poisson brackets between any two
potentials ,
$\{A_1 ,A_2\}_{qp}=\{A,B\}_{Q,P}\{Q,P\}_{pq}=0$,  the $CP^1$-extension of the
 $sdiff~\Sigma
$ Lie-algebra is $Abelianized$ ( no commutators) in the process.   DKS already
made the
 remark that their derivation was  also valid for
$U(1)$. To sum up, taking a $real$ slice  reduces  $C^4$-valued solutions to
$C^2$-valued ones ``killing'', in the process, the Poisson-brackets.

     The reader might feel unhappy with this fact. At the end of this section
we will discuss the other option
that happens when one does $ not$  take a real slice but, instead,
 imposes the $C^1$-valued
dimensional-reduction condition ( the complexification of eq-5c, below
) : $\partial_{x_1}-\partial_{x_3}=0$; where $x_1,x_3$ are $complex$
coordinates. Since in this case $Q^*$ is $ no$ longer equal to  $-P$, the
Poisson-bracket  is well defined, one  ends up still  having the
$CP^1$-extended $sdiff~\Sigma$ Lie-algebra untarnished  and with a
$C^3$-dependent  theory ( since the Plebanski equation was dependent of $C^4$).
Having $complex$-valued potentials  is precisely what is needed in order to
have the
$CP^1$-extended $sdiff~\Sigma$ to be locally isomorphic to $su^*(\infty)$.

Following  the same step by step procedure as the one outline below, yields
a $complexification$ of the KP equation. We just ask the reader to have
some patience to follow the steps below and later we  will come back to the
complexified-KP equation.

   The second step of the dimensional-reduction is to take
$\partial_{x_1}-\partial_{x_3 }=\partial_{x_{-}}=0$;
$x_{-}= x_1 -x_3.$  and,
hence, we end up with an effective $real$ three-dimensional theory.

Now we are ready to establish the correspondence with the dimensionally-reduced
$sl(2,R)$ SDYM
equations in $(3+3)$ dimensions by DKS. Set      :

$$x_1\rightarrow X_6=Y;~~x_2\rightarrow x_2(X,Y,T);~~x_3\rightarrow X_3=X;
{}~~x_4\rightarrow X_1=T. \eqno(5a)$$
$$\partial_{X_2}= 0.~~\partial_{X_3}=\partial_{X_4}.
{}~~\partial_{X_5}=0. \eqno(5b)$$
$$\partial_{x_{-}}=0 \Rightarrow u(X;Y;T) \rightarrow \Omega
(x_1+x_3;x_2;x_4)=\Omega (x_{+};x_2;x_4).\eqno (5c)$$
(Notice the variables in eq-(5c); $\Omega$ is a function of a spatial, timelike
and $null$ variable. Compare this with the variables in the KP
function ; two temporal  and one spacelike.)

 Notice that  $x_2$ cannot be mapped into a linear combination of $X_2,X_4,X_5$
because as a result of the DKS condition in (5b),
$\partial_{X_3}=\partial_{X_4}\Rightarrow
\partial_{x_3}=\partial_{x_2}=\partial_{x_+}.$  Such constraint is
incompatible with $\partial_{x_-}=0$ and the Jacobian :
$$     {\cal J}=\frac {\partial (X,Y,T)}{\partial (x_+,x_2,x_4)}=0
 \eqno (5e) $$

    Eq-(5a) defines a class of maps from ${\cal M}^{1+2}\rightarrow {\cal
N}^{1+1+1}$; i.e. $\phi : P(X,Y,T)\rightarrow P'(x_+,x_2,x_4)$. The
Jacobian $\cal J$ should not vanish and without loss of generality can be
set to one : ${\cal J}={\cal J}^{-1}=1 $. Hence, eq-(5a) defines a class of
volume-preserving-diffs.

     Using eqs-(5a-5c) in  equations (25,37,39,40,41) of DKS, we learn, from
the correspondence given in
eqs-(1,2), respectively, that a $one$ to $one$ correspondence with the
$SU(\infty)$  SDYM equations,
is possible iff we take for an Ansatz (see eq-46 in DKS)
$$A_{x_1}=A_{x_3}.\eqno (5d)$$
This ansatz is compatible with the dimensional reduction conditions in eq
-(5c) as we shall see below. The DKS-Plebanski 'dictionary' reads :
$$F_{36}=0\rightarrow F_{x_1 x_3}=F_{y\bar z}+F_{\bar y z}=0
(?)       . \eqno(6a) $$
$$F_{13} =0 \rightarrow F_{x_3 x_4}=iF_{z\bar z}=0 (?). \eqno(6b)$$.
$$F_{16}=0\rightarrow F_{x_1x_4}=F_{y\bar z}-F_{\bar y z}=0 (?). \eqno(6c)$$
The reason the right hand side of (6a) is zero is a $result$ of the
eqs-(5c,5d).

Exactly the same happens to eq-(6b) which is nothing but one of the  SDYM
equations. $F_{y\tilde  y}= \{\Omega_{\hat q},\Omega_{\hat p}\}= 0$ is a result
of the ansatz in
eqs-(3a-3d) and the  $2+2$ SDYM equations.
(This is not the case in the Euclidean regime). Eq-(6b) becomes then the
dim-reduced-Plebanski-equation, after
using the condition of (5c),  $\Omega_{z\hat q}-\Omega_{\tilde z p}=0$ (DRPE).
Equation-(6c) is zero iff (i). The ansatz of eq-(5d) is used. (ii). The
dim-reduction conditions in eq-(5c) are taken and, (iii). The DRPE  is
satisfied, eqs-(4,6b). If conditions (i-iii) are met it is straight
forward to verify that eq-(6c) =$(\partial_{x_1}-\partial_{x_3})A_{x_4}=0$.

The $crux$ of this work is to obtain the desired relationship between $u$ and
$\Omega$
in order to have self-consistent loop arguments and equations; to render
the right handsides of eq-(6a,6b,6c) to zero; and, to
finally, obtain the desired KP equation from the dimensional reduction of
the Plebanski equation (DRPE).

Having
estalished the suitable correspondence and the assurance that the right-
hand-sides of eqs-(6a,6b,6c) are in fact zero, we can now claim, by $cons
truction$, that
 equations-(39,40,41,47,53) of DKS are the equivalent, in the $u$ variable
language,
to eqs-(6a,6b,6c) above, in the $\Omega$ language.
 Therefore the equivalence for eq-(6c) reads :

$$ F_{x_1 x_4}=\Omega,_{
qq}+\Omega,_{pp}-\Omega,
_{q\bar z}-\Omega,_{pz} =  $$
$$1/2(\partial^2_y-\partial_y\partial_{\bar z}
+\partial_{\bar y}\partial_z +\partial^2_{\bar y})\Omega =
u_T-(1/4)u_{XXX}-(3/2)uu_{X}+(\lambda^2
+\alpha\beta)u_{X}=0.     \eqno (7a)$$
And, similarly, eqs-(47) of DKS are the equivalent of eq-(6a) above :

$$u_{Y}-\beta u_{X}=
1/2(\partial^2_y-\partial_y\partial_{\bar z}-\partial_{\bar y}\partial_z-
\partial^2_{\bar y})\Omega =0. \eqno (7b)$$

(Of course, one has to make a suitable scaling of the variables because the
KP equation obtained by DKS was given in terms of dimensionless quantities).
We must emphasize that $\Omega$ is not constrained, in any way whatsoever,
by satisfying two differential equations. The l.h.s of (7a) is zero as a
consequence of the DRPE; the condition $\partial_{x_{-}} =0$ and
$A_{x_1}=A_{x_3}$. Eq-(7a) is a $derived$ expression from the three latter
conditions.

After some tedious but straightforward algebra we can rewrite eqs-(7a,7b)
as follows :
$$u_{T}-1/4u_{XXX}-3/2uu_{X}+(\lambda^2+\alpha\beta)u_{X}=$$
$$1/2(\partial^2/\partial {x^2_+}-\partial^2/\partial {x^2_2}
-i\partial^2/\partial {x_+}\partial {x_4}
-i\partial^2/\partial {x_+}\partial {x_2})\Omega
(x_{+};x_2 ;x_4 ) =0.  \eqno (8a)      $$

$$\beta u_{X}-
u_{Y}=1/2(-\partial^2_{x_+} +\partial_{x_2}\partial_{x_4}+
2i\partial_{x_+}\partial_{x_2})\Omega =0.\eqno(8b)$$

and we include the DRPE  :

$$1/2(\partial^2_{x_+} + \partial_{x_2}\partial_{x_4})\Omega (x_{+};x_2;x_4)=0
\rightarrow (DRPE)\eqno (8c)$$

The function, $u (X,Y,T)$, satisfies the KP equation :

$$\partial_{X}(u_{T}-1/4~u_{XXX}-3/2~uu_{X})=-(\lambda^2+\alpha\beta)\beta^{-2}~u_{YY}.\eqno (9)
)$$
after  using the relation, $u_{Y}=\beta u_{X}\rightarrow
 u_{YY}=\beta^2 u_{XX}$ and differentiating the l.h.s. of (8a).

We still haven't finish yet; eqs-(8a,8b) are ``similar'' to the Backlund-type
transformations which
express  solutions of the DRPE, eq-(4), to solutions of the KP equation in
(9), by relating all first-order derivatives of $u$ to functionals  of
$\Omega,u$ and,  derivatives thereof. However, these
``Backlund-type''transformations are of no much use because both sides  of
eqs-(8a,8b) are zero; i.e. one ends with the tautology, $0=0$.

      The way to procced goes as follows. We have seven equations to deal
with. These are :

(i). The lhs and rhs of eqs (8a).

(ii). The lhs and rhs of eqs (8b)

(iii). The DRPE, eq-(8c)

(iv). The KP equation, (9).

(v). Equation (5e), nonvanishing Jacobian.

   We have all what is needed in order to  arrive finally to our main result of
this paper :
For every solution  $\Omega$ of the DRPE (8c) we set : $\Omega [x_+(X,Y,T);x_2
(X,Y,T);x_4(X,Y,T)]=u(X,Y,T)$ and plugging $\Omega$  into the $l.h.s$ of
(8a,8b) we
get three partial differential equations, once we include (5e) :
 ${\cal J}={\cal J}^{-1}= 1$, for the volume-preserving diffs, $\phi :
P(X,Y,T)\rightarrow P'(x_+,x_2,x_4)$. Once a solution for the three
diffeomorphisms
,that comprise $\phi$, is found then we have an explicit expression for
$u(X,Y,T)$ that solves the KP equation $by~construction$

         And, viceversa, once a solution for the KP equation is found, $u$,
 we set $u[X(x_+,x_2,x_4);Y(...);T(...)]$ equal
to $\Omega (x_+,x_2,x_4)$ and plugging
$u$ into the $r.h.s$ of (8a,8b) we get three partial differential equations
, once we include (5e)(nonvanishing Jacobian), for the inverse
volume-preserving-diffs, $\phi^{-1} :
P'(x_+,x_2,x_4)\rightarrow P(X,Y,T)$. Once a solution is found, we then
have an explicit expression for $\Omega (x_+,x_2,x_4)$ that solves the DRPE
$by~construction$ because eqs-(6a,6c) $\Rightarrow$ eq-(6b).

   In a separate publication we shall explain in more detail why this
construction that relates the KP to Plebanski is precisely what furnishes
the geometrical meaning of $W$ gravity. Remember what we said earlier about
the fact that the  DRPE provides a solution-space, $\cal S$,  of dim-reduced
hyper-Kahler metrics in  the
$complexified$ cotangent space of the Riemannian surface, $(T^*\Sigma)^c$,
required in the
formulation of $SU^*(\infty)$ $2+2$ SDYM theory. Since the
volume-preserving diffs, $\phi;\phi^{-1}$, yield the dim-reduced- Plebanski-
KP correspondence; it is natural that the $W_{1+\infty}$ symmetry algebra
associated with the KP equation $is$ the $\phi~transform $
of the $dimensional~reduced~CP^1$-extended $sdiff~\Sigma$- Lie algebra.

   Because these $\phi$, volume-preserving- diffs, act on the $ solution~
space, \cal S$, alluded earlier, it becomes clear that the $W~metric$
can be interpreted as the gauge field which gauges these $\phi$
diffeomorphisms acting on the space $\cal S$ !!
In [1] we already made the remark that one could generalized matters
evenfurther by starting with a $SU(\infty)$ SDYM theory on a
$self~dual~curved~2+2$
background. Upon imposing the ansatz of eqs-(3a-3d) one gets a
generalization of equation (4) where an extra field, $\Omega^1$, the
Plebanski first heavenly form , appears in addition to $\Omega$. $\Omega^1$
is the field which encodes the self-dual metric for the $D=2+2$ background.
The generalization of (4) encodes the interplay between ``two'' self-dual
``gravities''; one stemming from $\Omega$, the other from $\Omega^1$. In
any case, if we wish to gauge the volume-preserving-diffs, $\phi$, in order
to get $W$ gravity, where the $W$ metric is the $gauge$ field, we need to
come up with an extra field, which is the role played precisely by $\Omega^1$.

    What happens when we set the Jacobian, $\cal J$ to an arbitrary
constant, $\lambda$? In this case one gets a class of
volume-preserving-diffs depending on $\lambda,~\phi_{\lambda}$ which
furnishes the $W_{\infty}(\lambda)$ algebras discussed in the literature
[6]. A further and detailed discussion of this will be presented in a
future publication.

   In the case that one imposes a $truncation$ the algebra should reduce to
$W_N$. We still haven't decipher how to do that. Our construction and the
ones by Park [4] and Ablowitz and Chkravarty [10] rely on the $N\rightarrow
\infty$ limits, which is not unique to begin with. For finite $N$ the
geometrical picture is lost.

     We've been working with a dim-reduction procedure; is it possible to
incorporate other types of reduction schemes, like Killing symmetry types ?
 Park [4] showed  that a Killing symmetry reduction of the Plebanski first
heavenly equation yields the $sl(\infty)$ continual Toda equation whose
asymptotic expansion by Ablowitz and Chakravarty [10] led to the KP
equation.  The $W_{\infty}$ algebra was obtained as a Killing symmetry
reduction of the $CP^1$-extended $sdiff~\Sigma$-Lie algebra.  Since the physics
of self dual gravity  should be  independent
of which formulation one is using, either in terms of the first or second
heavenly form, this fact corroborates, once more,  that our findings should be
correct because
$sl(\infty,C)$ can be embedded into $sl(\infty,H)\sim su^*(\infty)$.
Gervais and Matsuo [17] proposed viewing $W$ gravity as holomorphic  embeddings
of $\Sigma$ into Kahler coset
spaces  $G/H$ where $W$ transformations look like diffs in $G/H$.
This would fit into our picture when a Killing-symmetry reduction scheme is
chosen. Furthermore, the role of instanton-like embeddings of $\Sigma$
into $G/H$ was noticed. It is clear that a lot remains to be done.

     We could have given, from the start, eqs-(8a,8b) ( without equating them a
priori to zero) as the prospective
$\Omega \rightarrow u $ transformations and,  after recurring to the
$SU(\infty)$
SDYM
equations, the DRPE, the anstaz that $A_{x_1}=A_{x_3}$, etc..... arrive
at the KP equation. However we would not have had any clues as to why and
where  all these equations stemmed from. In otherwords, one could have
anything on the l.h.s of (8a,8b)!. But this $cannot$ be the case because
the l.h.s of (8a, 8b) must have the form dictated by the $DKS$-$Plebanski$
correspondence.

    As mentioned earlier one could have $ not$ taken necessarily  a real slice
which
rendered the collapse of the Poisson brackets. Instead,  we could have taken
the complexification of eq-(5c),  simply by holding on to the initial
$complex$ valued, $x_1,x_2,x_3,x_4$ coordinates. Therefore, the
complexification of $\partial/ \partial x_{-} =0$ reduces $C^4$-dependent
solutions of (4) to
$C^3$-dependent ones and, after following, $mutatis~ mutandis$, the steps
hereby
presented, arrive at the complexified version of the KP equation. This is
not difficult to verify by simply keeping in mind the
 $ sl(2,R)~Self~Dual~DKS \rightarrow SU(\infty)~SDYM$ correspondence.
By a simple inspection of the $(2+2)$ SDYM equations and eqs-(3a-3d),  one
can see that the DRPE ( and all of
our equations) remains unchanged in the complexified case as well.
This would not be the true  in the Euclidean case; the Poisson-brackets do
not decouple in such case. The complexified KP equation has an advantage
over the real KP one,  since in the former,  the Poisson-brackets do not
collapse despite their  decoupling in the final  complexified DRPE, and the
$W_{\infty}$ symmetry would appear as a reduction of the $CP^1$-extended
$sdiff~\Sigma$ Lie-algebra [4]. ( Rigorously speaking, it is the
classical $W_{1+\infty}$ algebra which is related with the first Hamiltonian
structure of the KP hierarchy [6] ). In any case, we can see the
geometrical origin of these  classical $W_{\infty}$ algebras as
area-preserving diffeomorphisms.

  The KP equation has been obtained without any perturbation nor asymptotic
expansion methods : It has been obtained from $geometrical$ means
and from the crucial role that the $sdiff~\Sigma$ Lie-algebra, a bracket
preserving one, has on two-dimensional
 physics  ( strings can be obtained   upon dimensional
reduction of  membranes ). Evenfurther, the
importance  that the self-dual gravitational background, $(T^*\Sigma)^c$,
 and Plebanski's equation has  on the geometrical derivation of the KP equation
has been
unravelled. We have seen how self-duality in $(3+3)$
dimensions for the $SL(2,R)$ gauge group is intimately connected to
self-duality for $SU(\infty)$ Yang-Mills theory in $(2+2)$ dimensions.
And, finally we have seen how $W$ gravity can be constructed as the gauge
theory of the $\phi$ diffeomorphisms acting on the space, $\cal S,$ of
dim-reduced
$D=2+2~ SU^*(\infty)$ Yang-Mills instantons.

     Crucial for our derivation has been the fact that we were able to embed
$SL(2,R)$ into $SU(\infty)$. This is precisely what justified the
DKS-Plebanski correspondence ! One is essentially embedding
$R^8\sim C^4$-dependent  solutions into $C^6$-dependent  ones. The embedding of
$SL(2,R)$ into $SU(\infty)$ requires the addition of two-real internal
coordinates and, hence, the $SL(2,R)$ SD theory in six dimensions has $8$
real degrees of freedom because the $sl(2,R)$-valued  potentials ( $\infty
\times \infty$ matrices)  now  depend  on
two additional real variables ( representing the continous version of
discrete matrix indices) . In similar vein, the complexified $SU(\infty)$
SDYM theory in $C^4$ requires the addition of the two  complex-valued
canonical-conjugate coordinates, $Q(q,p),P(q,p)$, leaving us with
 $C^6$-dependent solutions . The latter coordinates  are just two
independent ( $Q\neq \lambda P;~\lambda = constant$)  maps
of the sphere, per example, to $C^1$. Since the sphere is topologically
$CP^1$, we have here a simple explanation of why the $CP^1$
extension of the $sdiff~\Sigma$ Lie-algebra $is$ the symmetry algebra of
Plebanski's equation; this was indeed  proven by Park [4] within
the framework of sheaf cohomology. To complete the actual counting we have
that the six $sl(2,R)$  potentials yield a total of $6\times 8 = 48$ real
degrees of freedom whereas the $su(\infty)$ ones yield :  $4\times 6 \times 2
= 48 $ real degrees of freedom.  This matching is another sign of
consistency that validates the $a~ priori$  $DKS$-$Plebanski$ correspondence .

 Recently, Nishino has shown that  the super-KP equation
can be embedded into  a self-duality
condition in $(2+2)$ superspace [14]. Using the results by us in [1] we
can automatically extend the construction here to the supersymmetric case.
In [1] the Lorentzian version of the Plebanski equation was derived for
$(3+1)$ superspace :
$$(\Theta,_{pq})^2 -\Theta,_{qq}\Theta,_{pp}+
\Theta,_{qz}-\Theta,_{p\bar z}=0.   \eqno (13)$$

where $\Theta$ is a light-cone chiral superfield described by Gilson et al
[15]. For self-duality conditions in Euclidean and Atiyah-Ward spacetimes,
spaces of signature $(2+2)$,  see [1].

     It is not difficult to see how this gauge-theoretical and geometrical
approach, can provide us with many important clues as to why these models
are integrable and, what is more important, its  higher-dimensional
origin.  The task now is to go to ten dimensions and use the power of
twistors methods  in higher dimensions to study the generalization of
self-dual theories in $D$ greater
than four [16].

      To conclude,  as far as we know,
 the physical derivation of the KP equation, besides
DKS and the references therein [7],  has been based on a
perturbation/asymptotic
expansion method : in  the original sea-waves  equation proposed by Stokes in
$1847$  the
weakly nonlinear, weakly dispersive, and weakly two-dimensional effects
were all of the same order [10]. Also, the one
based on the asymptotic $h\rightarrow 0$ limit of the continual $A_{\infty}$
Toda
molecule equation by Chakravarty and Ablowitz neglected powers of $h^6$ and
higher [10].

Eq-(9) was obtained $without$ any perturbation/asymptotic expansion method; it
is
$exact$ : it was solely based on a dimensional- reduction of the $SU(\infty)$
SDYM
equations in $(2+2)$ dimensions and on the correspondence, provided by
eqs-(1,2), into the DKS equations.

The weave  amongst  SDYM, self-dual gravity, topological field theories,
integrable models, $W$ algebras and  the   KP hierarchy seems to be getting
more and more tightly woven, especially in the supersymmetric case.
We believe, in view of the results presented above, and the fact
 that the  $N=2~SWZNW$ model valued in $sdiff~\Sigma$ is tantamount of
$4D~Self~Dual~Supergravity$ [3],  that $ supersymmetric$  $SU(\infty)$ SDYM
theories should generalize
Topological Field Theories. The recent work by the Trieste group
 on hyper-instantons and quaternions  already hints towards  more subtle
generalizations.
The fact that the quaternions appear in the isomorphism, $sl(N,H)\sim
su^*(2N)$ is very suggestive that we are on the right track.

\vspace{7 mm}

\centerline {REFERENCES}
\vspace{7 mm}
1. C. Castro : Journ. Math. Phys. 34, 2  (1993) 681.

2. E.G.Floratos, J. Iliopoulos, G.Tiktopoulos : Phys. Lett. B 217 (1989)
285.

3. C. Castro : ``The $N=2~ SWZNW$ model valued in $ sdiff~\Sigma$  is Self Dual

Supergravity in four Dimensions ``. I.A.E.C-11-92 preprint; submitted

 to the Journ. Math. Phys.

4. Q.H. Park : Phys. Letters. B 236 (1990) 429; ibid  B 238 (1990) 287.

``A 2D-sigma  model approach  to 4D instantons `` Int.Jornal of Modern Phys

A; page 1415, (1991).

5. I. Bakas : Nucl. Phys. B 302 (1988) 277.

   V.Drinfeld, V. Sokolov : J. Sov. Math. 30 (1985) 1975.

   A.Bilal, J.L.Gervais : Phys. Lett. B 206 (1988) 412.

   A. Das, S. Roy : J. Mod. Phys. 6 (1991) 1429

   J.L. Gervais : Phys. Lett. B 160 (1985) 277.

   F.Magri : Journ. Math. Phys. 19 (1978) 429.

6. P. Bouwknegt, K. Schouetens : ``W symmetry in Conformal Field Theories

`` CERN-TH-6583/92. To appear in Physics Reports.

7. A.Das, Z. Khviengia, E. Sezgin : Phys. Letters B 289 (1992) 347.

8. C. Castro : Phys. Lett. B 288 (1992) 291.

9. E.G.Floratos, G.K. Leontaris : Phys. Lett. B 223 (1989) 153.

10. M.J. Ablowitz, P.A. Clarkson : ``Solitons, Nonlinear Evolution

Equations and Inverse Scattering'' London Math. Soc. Lecture Notes 149;

Cambridge University Press (1991).

11. Edward Witten :''Surprises with Topological Field Theories ``

Proceedings of Strings' 90 at College Station, Texas, USA.

12. O. I. Bogoyavlenski : Comm. Math. Phys. 51 (1976) 201.

13. Jens Hoppe : Int. Journal Mod. Phys. A4 (1989) 5235.

14. H. Nishino : UMDPP-93-145-preprint :'' The super-KP systems embedded in

supersymmetric self-dual Yang-Mills theory''.

15. G.R.Gilson, I.Martin, A. Restuccia and J.G. Taylor : Comm. Math.

Physics 107 (1986) 377.

16. R. S. Ward, R.O. Wells : ``Twistor Geometry and Field Theory ``

Cambridge University Press. (1991).

17. J.L.Gervais, Y. Matsuo :'' Classical $A_n$ W geometry'' L.P.T.E.N.S
91/35 preprint.

\end{document}